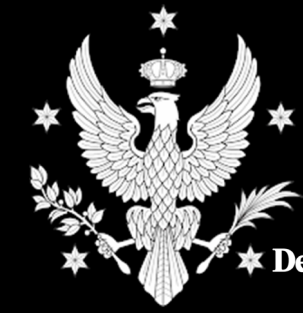
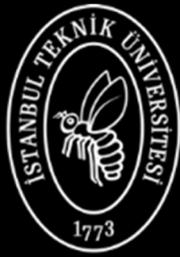

**Department of Complex System Modelling**
**Institute of Theoretical Physics**
**Faculty of Physics**
**University of Warsaw**

**Surface Treatment Group**
**Department of Metallurgical & Materials Engineering**
**Faculty of Chemical & Metallurgical Engineering**
**Istanbul Technical University**

# Ab-initio calculation of point defect equilibria during heat treatment

## Nitrogen, hydrogen and silicon doped diamond

### ( Preprint )

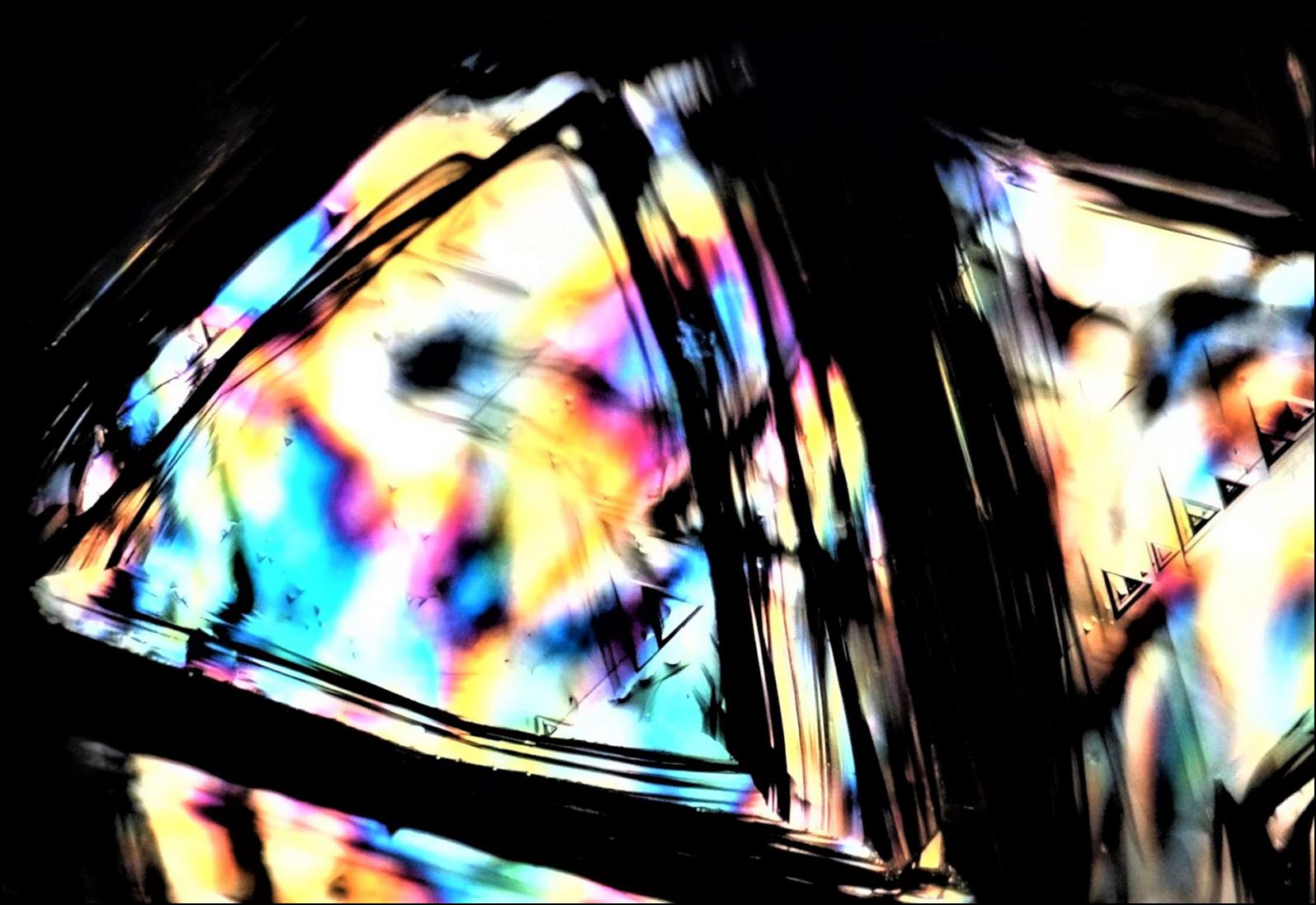

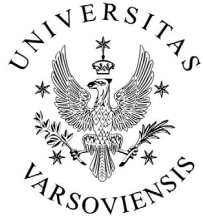 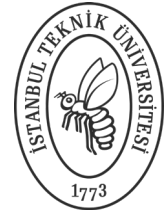

**İTÜ Surface Treatment Group**

**(Preprint)**

**November 22, 2021**

# Ab-initio calculation of point defect equilibria during heat treatment: Nitrogen, hydrogen, and silicon doped diamond

Mubashir Mansoor[1,2], Mehya Mansoor[1,3], Maryam Mansoor[1,4], Ammar Aksoy[1], Sinem Nergiz Seyhan[2], Betül Yildirim[1], Ahmet Tahiri[1], Nuri Solak[1], Kürşat Kazmanli[1], Zuhal Er[2,5], Kamil Czelej[6*], Mustafa Ürgen[1*]

[1] *Metallurgical and Materials Engineering Department, Istanbul Technical University, Istanbul, Turkey*
[2] *Department of Applied Physics, Istanbul Technical University, Istanbul, Turkey*
[3] *Geological Engineering Department, Istanbul Technical University, Istanbul, Turkey*
[4] *Mining Engineering Department, Istanbul Technical University, Istanbul, Turkey*
[5] *Maritime Faculty, Istanbul Technical University, Istanbul, Turkey*
[6] *Department of complex system modelling, Institute of Theoretical Physics, Faculty of Physics, University of Warsaw, Warszawa, Poland*

A B S T R A C T

Point defects are responsible for a wide range of optoelectronic properties in materials, making it crucial to engineer their concentrations for novel materials design. However, considering the plethora of defects in co-doped semiconducting and dielectric materials and the dependence of defect formation energies on heat treatment parameters, process design based on an experimental trial and error approach is not an efficient strategy. This makes it necessary to explore computational pathways for predicting defect equilibria during heat treatments. The accumulated experimental knowledge on defect transformations in diamond is unparalleled. Therefore, diamond is an excellent material for benchmarking computational approaches. By considering nitrogen, hydrogen, and silicon doped diamond as a model system, we have investigated the pressure dependence of defect formation energies and calculated the defect equilibria during heat treatment of diamond through ab-initio calculations. We have plotted monolithic-Kröger-Vink diagrams for various defects, representing defect concentrations based on process parameters, such as temperature and partial pressure of gases used during heat treatments of diamond. The method demonstrated predicts the majority of experimental data, such as nitrogen aggregation path leading towards the formation of the B center, annealing of the B, H3, N3, and $NVH_x$ centers at ultra high temperatures, the thermal stability of the SiV center, and temperature dependence of NV concentration. We demonstrate the possibility of designing heat treatments for a wide range of semiconducting and dielectric materials by using a relatively inexpensive yet robust first principles approach, significantly accelerating defect engineering and high-throughput novel materials design.

## 1. Introduction

As in all semiconducting materials, point defects in diamond are responsible for a wide range of optical and electronic phenomena [1]. Creating a defect of interest in sufficient concentration makes it possible to tailor diamond's optical and electronic properties for designing highly engineered optoelectronic devices [2]. The NV center is an excellent example of the significance of defects. Nitrogen-Vacancy center (NV) in diamond is a room temperature qubit with a long spin coherence time. The possibility to optically read out the NV center's spin states with present day technology can potentially revolutionize future electronics [3]. Therefore, the NV center is a prime choice for quantum computers [4]. The possibility of creating donor and acceptor defects in diamond, which is also corrosion resistant and biocompatible, has made it an attractive optical sensor for the bionic eye [5, 6]. Yang et al.

[7] and Koizumi et al. [8] have reviewed the electronic behavior induced by various defects in diamond. Color and luminescence are two additional examples where point defects play a crucial role [9, 10]. Trace amount of nitrogen dopant in approximately 1 ppm concentration is sufficient to completely transform diamond's optical and electronic behavior and cause a wide range of colors [11]. The induced coloration and luminescence depend on the type of point defect nitrogen atoms create. For example, a single carbon vacancy (V), known as GR1 center, is responsible for green color [12], while $N_3V$ defect (three substitutional nitrogen atoms surrounding a carbon vacancy, known in diamond literature as N3 center) is responsible for a yellowish color, which is a common point defect in natural diamond [13]. Some colors are produced only through the co-existence of several defects, as in the case of pink diamonds [14]. The comprehensive reviews by Zaitsev [15], Magna et al. [16, 17], and Haisnchwang et al. [18, 19] provide in-depth information on the cause of color in diamond.

Defects are an inherent part of materials as the configurational entropy favors their formation [20]. However, the incorporation of defects can be either intentional or unintentional. The former is a technological triumph, especially in the past few decades with the rise of the semiconductor industry [21, 22]. The latter is most common in natural and industrial processes, as seen in nitrogen, hydrogen, and silicon doped diamond. Nitrogen solubility in diamond is approximately 2000 ppm [23]. Therefore, it is ubiquitous in most mined natural diamond crystals and present in most laboratory grown crystals. It is possible to grow diamond at low pressures through chemical vapor deposition in either hot filament (HF-CVD) [24, 25] or microwave plasma (MW-CVD) systems [26, 27]. In both cases, the reactants are generally carbon carrying gases such as methane or acetylene and hydrogen gas, which form methyl radicals. $CH_3$ radicals are subsequently reduced to elemental carbon through further reactions with atomic hydrogen [28], although this is not the only reaction pathway in a CVD chamber [29]. The high efficiency of microwave plasma in dissociating the $H_2$ molecule [30] makes it a very suitable technique for growing optically high quality diamond crystals in relatively short time frames. One of the frontiers in diamond synthesis research is the growth rate. The addition of nitrogen gas into a CVD chamber, even in trace quantity, can increase the growth rate several times [31, 32]. Therefore, considering the relatively high solubility of nitrogen and hydrogen in diamond and their presence during crystal growth, N and H atoms are usually unintentional dopants in CVD grown diamond crystals. An additional unintentional dopant is silicon. Incorporation of Si occurs through etching of chamber sight glasses [33] and substrates [34] by hydrogen plasma, and therefore silicon related defects are widely reported in CVD grown diamond crystals [35]. Nitrogen, hydrogen, and silicon atoms can create a plethora of point defects in diamond, in combination with each other and intrinsic defects such as vacancy. However, not all defects are stable from a thermodynamic standpoint, and their concentrations can be highly dependent on the process parameters, such as temperature and partial pressures of gases involved [36]. Therefore, it is possible to transform defects through post-processing methods, such as heat treatment [37], to alter diamond's optical and electronic behavior.

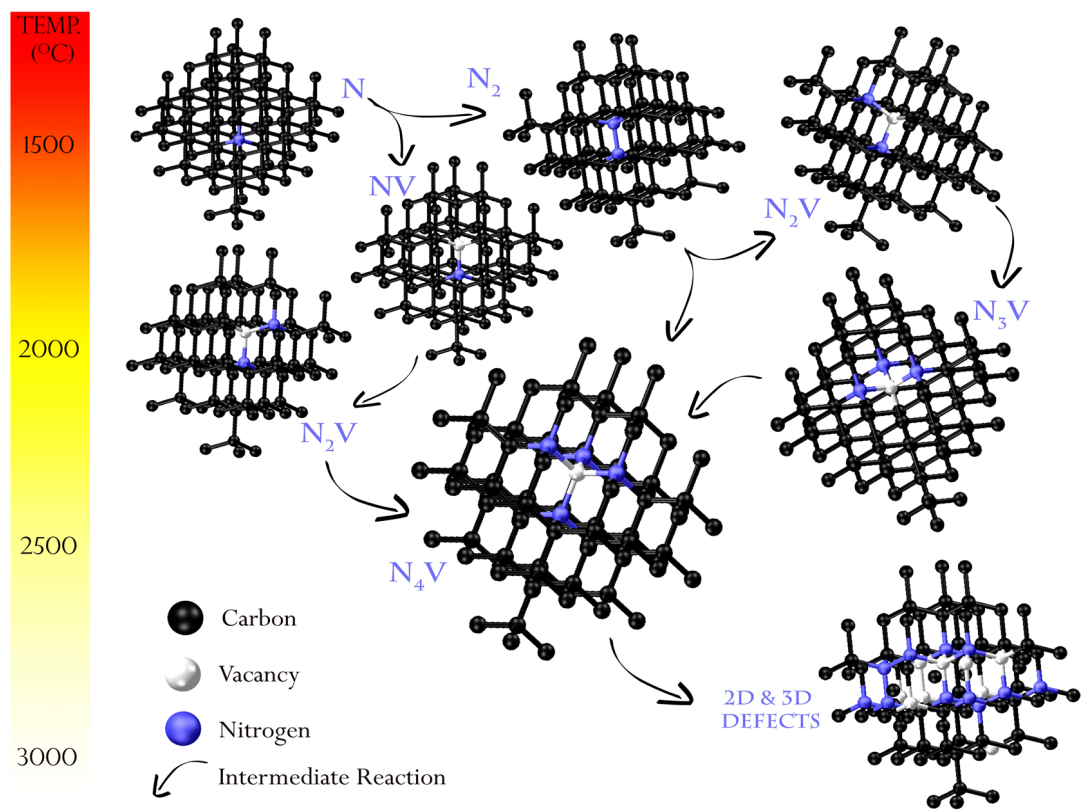

Fig. 1. A schematic representation of the nitrogen aggregation path reported experimentally [11, 37]. Single substitutional nitrogen (C center) tends to aggregate until $N_4V$ defect (B center) forms, with NV, $N_2V$ (H3 center) and $N_3V$ (N3 center) being intermediate reaction steps. There is clearly a thermodynamic drive towards the B center. The $N_4V$ and $N_3V$ defects are known to anneal out at ultra high temperatures, and extended 2D and 3D defects such as nitrogen platelets and voidites [51] are formed under such extreme conditions.

Given the significance of color on the value of a diamond in the gem and jewelry industry, a variety of heat treatments have emerged over the past few decades, which tend to transform point defects, thereby changing the color of diamond crystals [37, 38]. There are two main methods of heat treatment in diamond: low pressure high temperature (LPHT) and high pressure high temperature (HPHT). The former takes place under pressures of less than 1 atm; therefore, the duration of this treatment at higher temperatures should be limited to avoid graphitization of the entire crystal [39, 40], as diamond is not thermodynamically stable at pressures below 2.5 GPa [41]. The latter occurs at higher pressures where diamond is the stable phase of carbon; therefore, extended heat treatment durations become possible. Pre-treatments by various beams, such as electron irradiation, are common practices that increase vacancies by order of magnitude and thus accelerate diffusion kinetics [42]. Kazuchits et al. [43] have conducted a comprehensive analysis of LPHT and HPHT methods and have found that the defect equilibria are qualitatively similar in both cases. Through controlled heat treatment procedures, it is possible to transform defects extensively and produce yellow, pink, blue, red, and even colorless diamonds simply by adjusting the defect equilibria, which the dopants can create, without introducing any new dopants into the crystal [37]. These transformations are achieved through defect-defect reactions [44, 45] or equilibration of chemical potentials with the surrounding environment [46]. The vast pool of accumulated literature on color transformations through heat treatments in diamond makes this material an excellent choice for modeling defect thermodynamics. The detailed experimental literature on diamond allows benchmarking a computational approach for estimating defect equilibria during heat treatments. Such a luxury does not exist in most materials.

Some of the most commonly observed defects in as-grown CVD diamond are $N_s$ (single substitutional nitrogen), followed by $NVH_x$ defects (nitrogen-vacancy with hydrogen atoms passivating carbon dangling bonds) [47]. Feng et al. [48] have elegantly demonstrated the reason behind the abundance of these defects during crystal growth through defect formation energy calculations on hydrogen terminated diamond surfaces. However, when defect formation energies in the bulk crystal are considered, nitrogen aggregation should commence, as $N_2$, $N_3V$, and ultimately $N_4V$ have lower formation energy than $N_s$. Computational studies [49, 50] on bulk materials have demonstrated this phenomenon. In other words, during diamond synthesis, the thermodynamic equilibria favor the formation of $N_s$ and $NVH_x$ defects. However, once these defects are locked within the bulk diamond and the crystal has grown further, the thermodynamic defect equilibria change, favoring the formation of aggregated nitrogen related defect-complexes. The nitrogen aggregation process is widely reported during heat treatment experiments of diamond [37, 51], and it is also seen in diamond crystals which are naturally subjected to high pressure and high temperature for geological time scales [52, 53]. A comprehensive understanding of all intermediate aggregation steps does not exist; however, heat treatments have made it clear that the aggregation begins from the C center ($N_s$ defect) and proceeds towards the B center ($N_4V$ defect). NV, H3 ($N_2V$), and N3 ($N_3V$) centers are intermediate defect reaction products in the process [37]. The $N_3V$ and $N_4V$ defects are known to anneal out (diminish in concentration) at ultra-high temperatures over 2000 °C, and the appearance of voidites and nitrogen platelet line defects are reported at such extreme conditions [54, 55]. Fig. 1 presents a schematic illustration of the experimentally reported nitrogen aggregation path. Experimental studies by Evans et al. [56] and Taylor et al. [57] have attributed the kinetics of nitrogen aggregation in diamond to temperature and time factors. They have also derived an Arrhenius relationship between the percentages of B center ($N_4V$ defect) to other defects as an indication of the residence time of a diamond at a particular temperature. This relationship between the percentage of $N_4V$ defect with time and temperature is also used for an approximate dating of a diamond's genesis and formation temperature-time history [58] through spectroscopic intensities of the B center's spectra compared to other less aggregated defects. Similarly, mantle thermometry techniques have been developed based on nitrogen platelet defects in diamond [59].

The heat treatments of diamond in the gem and jewelry industry are mesmerizing examples of the power which post processing of crystals can have in defect engineering for solid state lasers, and advanced optoelectronic devices. However, the plethora of defects, which can co-exist, makes it challenging to map out their stability regions as a function of heat treatment parameters by experimental means. This complexity also makes it difficult to judge which defects warrant detailed computational insight without empirical data on defect equilibria. The existing experimental data on nitrogen aggregation in diamond alone has been gathered in over seven decades of research. Thus, accumulating similar data for other defect systems is not a trivial matter and requires years of experimental devotion to mapping out stable defects as a function of process parameters. On the other hand, prior ab-initio calculations [49, 50] on defect equilibria in diamond have relied on empirical adjustments of chemical potentials or equilibrium Fermi energy to reproduce experimental data, which is not a possible strategy for novel co-doped systems where experimental information does not exist. Therefore, it is essential to study defect thermodynamics from a fully ab-initio perspective and eliminate empirical input to achieve accelerated progress in defect engineering of novel dopant systems in materials.

Formation energy versus Fermi energy diagrams are widely reported for various defects in diamond and other materials. However, such diagrams are not meaningful at first sight from a process engineering perspective. Kröger-Vink diagrams, on the other hand, can make practical defect engineering through heat treatments easier. Kröger-Vink diagrams are widely used for illustrating defect concentrations at a constant temperature as a function of the partial pressure of oxygen or nitrogen for oxides and nitrides, respectively. By calculating defect concentrations for a wide temperature range, based on a canonical ensemble, and plotting what we call monolithic-Kröger-Vink diagrams to illustrate defect concentrations with respect to heat treatment temperature, it becomes possible to engineer heat treatments in an accelerated manner.

This study aims to plot monolithic Kröger-Vink diagrams for illustrating defect concentrations based on process parameters (temperature and partial pressures of gases) during heat treatments of N-H-Si doped diamond. Based on the prior thermodynamic models by Yildiz [60, 61] and Van de Walle [46, 62] research groups and ab-initio methods of point defect calculations [63], we have conducted a first principles investigation on the defect equilibria for nitrogen, hydrogen, and silicon doped diamond, as the vast experimental data on heat treatments of diamond allows effective benchmarking. The inexpensive yet robust modeling approach applied in this study can be adapted to most semiconducting and dielectric materials, which allows the creation of large databases for a plethora of defects. We demonstrate the possibility of defect engineering on novel co-doped materials through fully ab-initio modeling of heat treatments by considering a variety of defects.

## 2. Computational method

### 2.1. DFT inputs and defect formation energy

We have applied spin polarized density functional theory (SP-DFT) under periodic boundary conditions for calculating formation energies of 300 defects in diamond, which are 60 distinct combinations of substitutional nitrogen and silicon, carbon-vacancies, and interstitial hydrogen atoms, in five different charge states. Table 2 demonstrates a complete list of the considered defects. We used generalized gradient approximation as the exchange functional as parametrized by Perdew-Burke-Ernzernhof (GGA-PBE) [64]. The calculations have been carried out using Projector Augmented Wave (PAW) method [65], as implemented in Vienna Ab-initio Simulation Package (VASP 6.1) [66, 67], under the framework of MedeA 3.2 [68]. Defects have been modeled under reciprocal projection space using a 3×3×3 cubic supercell of 216 atoms (a = 3.57 Å), with spacing between k-points being 0.8 Å$^{-1}$ (Γ-only), cut-off energy of 500 eV, and Gaussian smearing of 0.05 eV. The self-consistency convergence (SCF) threshold of 10$^{-5}$ eV has been used, and the atoms were relaxed under a constant volume until Hellmann-Feynman forces were below 1 meV/Å. The defect formation energies ($\Delta H_f^q$) are calculated using Eq.1, as proposed by Van de Walle [46] and Zhang and Northrup [69].

$$\Delta H_f^{DFT} = E_{tot}^q - \sum n_i \mu_i + q(E_f + E_{VBM}) + E_{corr}^{FNV} \quad \text{(Eq. 1)}$$

$E_{tot}^q$ is the difference in the DFT calculated supercell energies of defective and pristine cells, $\mu$ and $n$ are the chemical potential and stoichiometric coefficient of added or removed elements in the defective cell, respectively. $E_f$ is the equilibrium Fermi energy of the crystal. The charge of a defect is given by $q$, and $E_{VBM}$ is the potential of the valence band maximum (VBM) in a pristine cell, as calculated using semilocal functional (GGA-PBE). $E_{corr}$ is added to correct for the finite supercell size and false electrostatic interactions due to the jellium background charge created during calculations of a charged defect, which obtains charge neutrality in the supercell [70]. Lany [71] and Freysoldt et al. [72] have presented a detailed treatment of this issue. We have applied the fully ab-initio correction as proposed by Freysoldt, Neugebauer, and Van de Walle (FNV) [73], through SXDEFECTALIGN code [74].

A shortcoming of the GGA-PBE functional is its underestimation of the band gap, which inevitably affects defect formation energies. However, Alkauskas et al. [75] have shown that defect levels as calculated by semilocal functionals can be used for approximating their hybrid counterparts by using a common reference potential, which can provide an estimate for the hybrid transitions within a reasonable level of accuracy. Defect thermodynamics does not require precise knowledge of defect transition energies because of the significant variations in equilibrium Fermi energy as a function of temperature (Appendix A). Therefore, applying a correction to the semilocal results to estimate the nonlocal energies of hybrid functionals poses the significant advantage of decreasing the necessary computational time by more than one order of magnitude, and yet reproducing defect equilibria in qualitative agreement with the hybrid functionals. We have applied the correction scheme proposed by West, Sun, and Zhang [76], as shown in Eq. 2.

$$\Delta H_f^{HSE} \cong \left[\Delta H_f^{DFT}(E_{VBM}^{DFT}) - q(\Delta \emptyset)\right] + q E_f^{HSE} \quad \text{(Eq. 2)}$$

By calculating $\Delta V$, which corresponds to the difference in vacuum potential with respect to bulk as calculated by GGA-PBE and HSE06 functionals (Fig. 2), we have found the correction factor $\Delta \emptyset$ as 0.478 eV.

The calculation of defect formation energies based on the method used requires a computational time of approximately 20 core hours per defect, which makes it possible to create comprehensive databases on a considerable variety of defects. The relatively low computational cost of this method allows pinpointing defects, which warrant further study by higher levels of theory through calculations of defect equilibria. Benchmarking the formation energies obtained in this study with previous HSE06 [77] calculations [49, 78-80] gave satisfactory results for estimating defect equilibria. The comparison between our data and prior calculations is available in Appendix B.

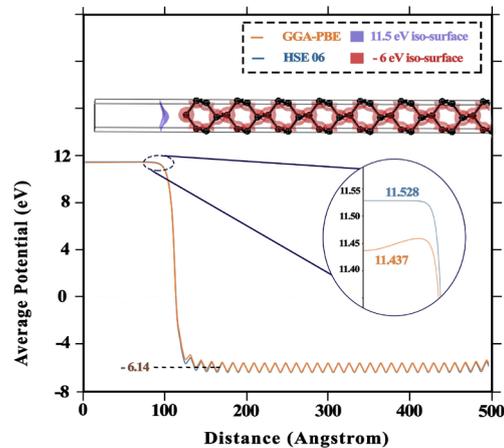

Fig. 2. The difference between vacuum and bulk potential of pristine diamond, as calculated by GGA-PBE (orange) and HSE06 (blue) functionals have yielded a correction factor of 0.478 eV. The correction factor is estimated by using the methodology proposed by West, Sun and Zhang [76].

The formation energy of a charged defect depends on the equilibrium Fermi energy of a crystal (Eq.1). $E_f$ can be considered as the electron chemical potential, a reservoir through which a canonical ensemble is formed where the equilibrium is only achieved through charge neutrality of the entire system [63, 81]. Therefore, $E_f$ depends on the concentration of all available defects in the solid. As elegantly stated by Freystoldt et al., [63], the benefit of this canonical approach is that no bookkeeping is necessary for individual reactions between defects. This was the difficulty with defect physics of Kröger and Vink [44] in the 1960s, and therefore, such a canonical approach simplifies the problem and increases the accuracy considerably since all possible defect reactions are taken into account. To compute $E_f$, Eq. 3, 4, and 5 should be solved in a self-consistent manner.

$$\sum q_i[d]_i + [p] - [n] = 0 \quad \text{(Eq. 3)}$$

Eq. 3 represents the boundary conditions for charge neutrality, where $[d]$ is the concentration of a defect, and $q$ is the defect charge. Hole and electron concentrations are given by $[p]$ and $[n]$, respectively, which have been computed using Eq. 4 and Eq. 5 [60].

$$[p] = \int_{E_{VBmin}}^{E_{VBmax}} \frac{g_v(E)dE}{1+\exp\left(\frac{E_f-E}{kT}\right)} \quad \text{(Eq. 4)}$$

$$[n] = \int_{E_{CBmin}}^{E_{CBmax}} \frac{g_c(E)dE}{1+\exp\left(\frac{E-E_f}{kT}\right)} \quad \text{(Eq. 5)}$$

Here, $g_v(E)$ and $g_c(E)$ represent the density of states for valence and conduction bands, respectively, as calculated by GGA-PBE for the pristine supercell. We have shifted the conduction band minimum to match the experimentally reported band gap of 5.47 eV to estimate electron and hole concentrations. In order to solve for $E_f$, an additional equation is necessary for calculating defect concentrations $[d]$, which can be evaluated using two different approaches: non-adiabatic and adiabatic. In the former approach, equilibration with a chemical reservoir ($\mu_i$) of the relevant species in the surrounding environment is assumed, making it a suitable approach for fast diffusing dopants and intrinsic defects. Such equilibration is expected for all dopants during the heat treatments of thin films or during crystal growth [46, 69], where the necessary diffusion distances for equilibration are relatively short.

However, when considering heat treatment of bulk materials, especially for crystals such as diamond where the diffusion coefficient of most atoms, except for hydrogen [82], are extremely low, equilibration of dopant chemical potential with the surrounding environment can occur either close to the surface, or if the crystal is exposed to high temperatures for geological time-scales. Therefore, modeling defects in bulk material during heat treatment processes requires a different approach, irrespective of the chemical potential of the dopant in the environmental reservoir. The calculations under the adiabatic constraints fulfill this requirement, in which the chemical potentials are computed using thermodynamic activities of the species involved within the solid solution and binding energies of complexes that can form [60]. Accordingly, we have used the adiabatic approach for dopants with a low diffusion coefficient (nitrogen and silicon) and the non-adiabatic approach for elements where equilibrations of chemical potentials are likely (hydrogen and carbon-vacancy). An overview of both methods is given in the following sub-sections.

### 2.2.1. Non-adiabatic approach

When considering the equilibration of a dopant atom's chemical potential with respect to chemical species in the surrounding environment, Eq. 6 is used. We have taken $H_2$ as the reference molecule of hydrogen and diamond for carbon chemical potentials. The incorporation of temperature into the defect formation energy is brought about through the chemical potentials of the reference molecules as implemented by Van de Walle [46].

$$\mu(T,P) = E_{ref} + \mu^0(T) + kT\ln(a) \quad \text{(Eq. 6)}$$

The $E_{ref}$ term is the DFT calculated energy of the reference atom described above, and $\mu^0$ is the change in chemical potential with respect to temperature, obtained from the thermochemical tables, such as the JANAF database [83]. The last term accounts for the thermodynamic activities, and for gases such as $H_2$, it is simply the partial pressure of the gas. By substituting Eq. 6 into $\mu_i$ in Eq.1, and considering defect concentrations $[d]$ as given in Eq.7, the equilibrium Fermi energy and defect concentrations can be solved self-consistently. It should be noted that under the non-adiabatic approach, the chemical potential is a direct indicator of a dopant's solubility in material, based on Eq.1, 6, and 7.

$$[d] = N_f N_c \exp\left(\frac{-\Delta H_f^q}{kT}\right) \quad \text{(Eq. 7)}$$

$N_f$ represents the number of possible defect configurations of the same energy, and $N_c$ is the number of maximum possible sites in diamond supercell for the defect [46, 63]. For simplicity, we have taken $N_f=1$ and $N_c$ as the number of carbon sites per unit volume for all defects. As seen in the proceeding sections, this simplification is justified based on their negligible consequence on the relative defect concentrations. The error that this simplification cause is inconsequential when considering orders of magnitude changes in equilibrium defect concentrations due to wide variations in equilibrium Fermi energy as a function of temperature.

### 2.2.2. Adiabatic approach

Considering dopants in the solid solution that cannot equilibrate with the surrounding environment, due to kinetic restrictions imposed during heat treatments, the defect-defect reactions are considered to dominate. In this approach, the total concentration of a doped element is assumed constant, and therefore, the reactions should satisfy charge neutrality and stoichiometric equilibrium for the entire system. In other words, each defect-complex is considered to be a product of its constituent defects, as shown in Eq. 8, where α, β, γ, and δ represent the number of silicon, nitrogen, vacancy, and hydrogen in the defect complex, respectively.

$$\alpha Si + \beta N + \gamma V + \delta H = Si_\alpha N_\beta V_\gamma H_\delta \quad \text{(Eq. 8)}$$

Under this approach, chemical potentials are calculated using Eq. 9. The activity is taken as the concentration of a reference defect. In other words, activities are variables in this case, unlike the non-adiabatic approach where activities are constant. $E_{def}$ and $E_{pure}$ represent the DFT calculated energy of reference defects and pristine cell.

$$\mu(T) = E_{def} - E_{pure} + kT\ln(a) \quad \text{(Eq. 9)}$$

The binding energies of the defects are used for evaluating concentrations [60], which are defined as follows. Please note that we have used the classical nomenclature where negative binding energy is exothermic and that the $E_i$ terms in Eq. 10 are DFT calculated energies of defective minus pristine supercells, which includes the FNV correction.

$$\Delta H_b^q = E_{Si_\alpha N_\beta V_\gamma H_\delta}^q - \alpha E_{Si} - \beta E_N - \gamma E_V - \delta E_H - kT\ln((a_{Si})^\alpha (a_N)^\beta (a_V)^\gamma (a_H)^\delta) + q(E_f) \quad \text{(Eq. 10)}$$

The activities for each constituent defect and equilibrium Fermi energy are unknown in Eq.10 and depend on defect concentrations, temperature, and total dopant concentrations. The concentration of each defect is given by Eq. 11.

$$[Si_\alpha N_\beta V_\gamma H_\delta]^q = N_c (a_{Si})^\alpha (a_N)^\beta (a_V)^\gamma (a_H)^\delta$$
$$* \exp\left(\frac{-\left(E_{Si_\alpha N_\beta V_\gamma H_\delta}^q - \alpha E_{Si} - \beta E_N - \gamma E_V - \delta E_H + q(E_f)\right)}{kT}\right) \quad \text{(Eq. 11)}$$

To solve for the unknowns and compute defect concentrations, charge neutrality as given in Eq. 3 and the following stoichiometric constraints are used, which are based on constant dopant concentrations at all times. In other words, the activities and equilibrium Fermi energy should be such that the total charge of a system is zero, and the total adiabatic element content in defects is equivalent to the total concentration of the dopant.

$$[Si]_{ppm} = \sum \alpha [Si_\alpha N_\beta V_\gamma H_\delta]_i \quad \text{(Eq. 12)}$$

$$[N]_{ppm} = \sum \beta [Si_\alpha N_\beta V_\gamma H_\delta]_i \quad \text{(Eq. 13)}$$

By solving the non-linear system of 300 term equations (as all defects in table 2 should be considered simultaneously when considering N, H, and Si doped diamond) numerically through the Newton-Rhapson algorithm [84], we have plotted monolithic-Kröger-Vink diagrams for defect concentrations as a function of temperature, process parameters and trace chemistry of a diamond. It should be noted that there are no kinetic considerations made in this paper, and we have calculated the thermodynamic equilibria only.

### 2.3. Pressure dependence of formation energies

We have investigated the pressure dependence of defect equilibria and the applicability of this study for heat treatments conducted under higher pressures (5-10 GPa), as in the HPHT process and super-deep natural diamonds, which are subjected to pressures over 40 GPa [85]. For this purpose, we have applied a hydrostatic pressure ranging from 0 to 50 GPa on a fully relaxed pristine cell, using the GGA-PBE functional as mentioned above. Defect formation energies of four neutral defects (N$_4$V, N$_3$V, N$_2$, and NV) were then calculated under constant supercell volume for each pressure, based on the inputs given in section 2.1. The nitrogen chemical potential for non-adiabatic condition is based on half of the formation energy of the N$_2$ molecule, as calculated using DFT under similar hydrostatic pressure as the diamond supercell. The carbon chemical potential is retrieved from the formation energy of diamond supercell per carbon atom. The formation energy under the non-adiabatic condition is calculated using Eq.1, and the binding energies are calculated using Eq. 10.

### 3. Results and discussions

#### 3.1. Pressure dependence of defect equilibria

Our calculations show that if nitrogen is considered an adiabatic specie locked within the diamond lattice, the ratio of binding energies for different defect complexes does not change considerably with respect to pressure, as shown in Fig. 3a. Given that the defect concentrations under adiabatic conditions are dictated by their binding energies, it can be deduced that the defect equilibria during heat treatment do not depend on pressure, mainly because heat treatment time scales cannot allow the equilibration of nitrogen chemical potential with the surrounding environmental nitrogen reservoir. This result is in direct agreement with the experimentally reported observations regarding the similarity of defect equilibria under HPHT and LPHT conditions and proves the hypothesis proposed by Kazuchits et al. [43] in this regard.

Table 1. Changes in carbon and nitrogen chemical potentials, lattice constant, density, and the change in band gap (referenced to zero GPa gap), with respect to pressure. Nitrogen chemical potential is taken as half the formation energy of the nitrogen molecule.

| Pressure (GPa) | μ$_C$ (eV/atom) | μ$_N$ (eV/atom) | Lattice Cons. (Å) | Density (g/cm$^3$) | ΔE$_g$ (eV) |
|---|---|---|---|---|---|
| 0.00 | -9.093 | -8.302 | 3.574 | 3.496 | 0 |
| 2.50 | -9.098 | -8.252 | 3.567 | 3.517 | +0.012 |
| 5.00 | -9.092 | -8.219 | 3.560 | 3.536 | +0.024 |
| 7.50 | -9.091 | -8.175 | 3.554 | 3.556 | +0.037 |
| 10.00 | -9.089 | -8.129 | 3.547 | 3.575 | +0.048 |
| 15.00 | -9.085 | -8.025 | 3.535 | 3.612 | +0.072 |
| 25.00 | -9.071 | -7.859 | 3.512 | 3.684 | +0.116 |
| 35.00 | -9.052 | -7.684 | 3.490 | 3.753 | +0.159 |
| 45.00 | -9.030 | -7.581 | 3.470 | 3.819 | +0.199 |
| 55.00 | -9.003 | -7.384 | 3.451 | 3.882 | +0.237 |

Considering nitrogen as a non-adiabatic specie, where only geologic time scales can allow the equilibration of nitrogen chemical potential with the surrounding nitrogen reservoir (referenced to $N_2$ molecule), the ratios between defect formation energies change substantially as pressure increases. Therefore, the defect equilibria under geological conditions are pressure dependent (Fig. 3b). The trend seen in Fig. 3b is primarily due to the increase in $\mu_N$ at higher pressures (Table 1). The reduction in the formation energy of $N_4V$ at higher pressure under non-adiabatic conditions dictates the inevitable higher concentration of the B center as pressure rises, even at zero Kelvin. The abundance of $N_4V$ defect seen in natural diamonds of super-deep origin [53, 86] formed in the mantle transition zone [85], where pressure is an order of magnitude higher than HPHT conditions, agrees with this calculated pressure dependent formation energy. Moreover, Eq. 11 clarifies that this reduction in the formation energy of the B center will be even more drastic as temperature increases. Therefore, minute traces of nitrogen gas can readily induce a high concentration of the $N_4V$ defect in diamond crystals under such geological conditions. In other words, HPHT treatment is not a true representative of the geological conditions that a natural diamond is subjected to, even if the pressure and temperature in both cases are the same.

Based on the results presented in Fig. 3a, we can conclude that the data and analysis provided in the following sections are applicable for heat treatments under LPHT and HPHT conditions. On the other hand, considering the results presented in Fig. 3b, the Arrhenius relationship proposed by Evans et al. [56] on the kinetics of nitrogen aggregation does not apply to the geological conditions in which natural diamond crystals are exposed. Therefore, applying the kinetics of nitrogen aggregation during HPHT treatments in diamond for geological dating purposes should be reconsidered to include the effect of pressure and geological time scales.

### 3.2. Defect equilibria

Based on the calculated DFT results given in Table 2, we have plotted monolithic Kröger-Vink diagrams as a function of temperature for several trace chemistries comprised of N, Si, and H dopants. Diamond crystals with 0.001, 1, and 1000 ppm of nitrogen content are considered. The former are characteristics of CVD grown crystals, and the latter is not rare in natural diamonds. Co-doping with silicon and hydrogen have also been considered, with Si taken as 1 ppm, and hydrogen has been considered to equilibrate with the surrounding atmosphere, based on the partial pressure of $H_2$ gas during heat treatment. We have considered two cases for $pH_2$, as $10^{-6}$ and 1 atm, which represent hydrogen poor and rich conditions during heat treatments.

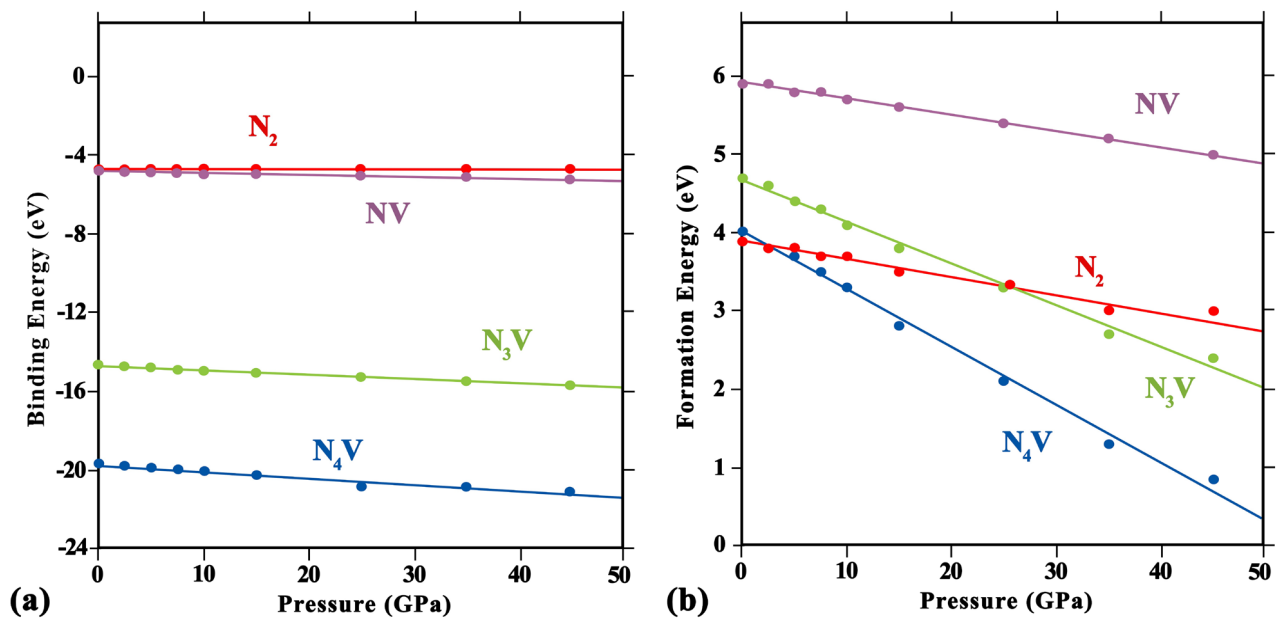

Fig. 3. Binding energies (a) and formation energies (b) of $N_2$, NV, $N_3V$, and $N_4V$ defects complexes are shown with respect to pressure for zero Kelvin temperature. The binding energies are almost independent of pressure, while the formation energies become highly pressure dependent due to the equilibration of nitrogen chemical potential with the surrounding environment. These results demonstrate pressure-independence of defect equilibria during heat treatments but pressure-dependence of equilibria under geological time scales.

Mansoor et al. (2021) / İTÜ Surface Treatment Group

Table 2. Point defects considered in this study are listed in this table. The calculated total supercell energies with the FNV charge correction are shown in the left half ($E_{tot} + E_{corr}$), and the right half presents the defect formation energies at zero Kelvin, considering an equilibrium Fermi energy of 3 eV. Chemical potentials $\mu_N$ is referenced to $N_2$ molecule and $\mu_{Si}$ to bulk silicon. Charged defects, which do not possess a defect level in the band gap are removed. All units are in eV.

| | $E_{tot} + E_{corr}$ | | | | | $\Delta H_f$ (0 K, 3 eV) | | | | |
|---|---|---|---|---|---|---|---|---|---|---|
| Charge / Defects | -2 | -1 | 0 | 1 | 2 | -2 | -1 | 0 | 1 | 2 |
| N | 33.166 | 18.743 | 5.124 | -7.860 | -16.925 | - | 5.688 | 4.349 | 3.645 | - |
| $N_2$ | 33.835 | 19.447 | 5.498 | -4.931 | -13.787 | 7.725 | 5.617 | 3.948 | 5.799 | - |
| $N_3$ | 38.949 | 22.308 | 9.894 | -1.656 | -11.041 | - | 7.703 | 7.569 | 8.299 | - |
| $N_4$ | 39.659 | 25.749 | 14.265 | 2.602 | -8.280 | 11.999 | 10.369 | 11.165 | 11.782 | 13.180 |
| V | 40.654 | 26.920 | 15.470 | 4.889 | -5.043 | 7.002 | 5.548 | 6.378 | 8.077 | 10.425 |
| NV | 41.327 | 27.459 | 15.772 | 5.022 | -4.282 | 6.900 | 5.312 | 5.905 | 7.435 | - |
| $N_2V$ | 42.129 | 28.066 | 16.102 | 5.622 | -3.868 | 6.927 | 5.144 | 5.460 | 7.260 | - |
| $N_3V$ | 43.321 | 28.847 | 16.134 | 5.832 | -3.376 | 7.344 | 5.150 | 4.717 | 6.695 | - |
| $V_2$ | 50.931 | 38.553 | 26.805 | 16.394 | 6.658 | 8.187 | 8.089 | 8.621 | 10.490 | 13.034 |
| $NV_2$ | 51.655 | 39.049 | 27.165 | 16.626 | 7.355 | 8.136 | 7.810 | 8.206 | 9.947 | 12.956 |
| $N_2V_2$ | 52.250 | 39.278 | 27.578 | 17.365 | 7.809 | 7.956 | 7.264 | 7.844 | 9.911 | 12.635 |
| $V_3$ | 64.774 | 52.335 | 40.696 | 31.242 | 20.629 | 12.938 | 12.779 | 13.420 | - | 17.913 |
| $NV_3$ | 68.415 | 56.062 | 44.355 | 33.854 | 24.224 | 15.804 | 15.731 | 16.304 | 18.083 | 20.733 |
| H | 30.276 | 15.775 | 2.722 | -9.022 | -18.113 | - | 6.880 | 6.107 | 6.643 | - |
| NH | 31.721 | 17.915 | 5.887 | -5.270 | -15.214 | 9.771 | 8.245 | 8.497 | 9.620 | 11.956 |
| VH | 37.085 | 24.031 | 11.960 | 1.849 | -7.689 | 6.818 | 6.044 | 6.253 | 8.422 | 11.164 |
| NVH | 36.497 | 22.891 | 11.061 | 1.110 | -8.297 | 5.455 | 4.129 | 4.579 | 6.908 | 9.781 |
| $N_2VH$ | 38.121 | 23.643 | 11.290 | 1.159 | -8.016 | - | 4.106 | 4.033 | 6.182 | - |
| $V_2H$ | 46.558 | 34.013 | 22.429 | 12.014 | 2.794 | 7.199 | 6.934 | 7.630 | 9.495 | - |
| $NV_2H$ | 47.235 | 34.247 | 22.755 | 12.073 | 2.778 | 7.101 | 6.393 | 7.181 | 8.779 | 11.764 |
| $V_3H$ | 60.393 | 48.019 | 36.481 | 26.223 | 16.356 | 11.942 | 11.848 | 12.590 | 14.612 | 17.025 |
| $H_2$ | 29.770 | 15.319 | 1.306 | -8.314 | -17.522 | - | 9.809 | 8.076 | 10.736 | - |
| $NH_2$ | 29.491 | 14.968 | 3.778 | -8.358 | -17.478 | - | 8.683 | - | 9.917 | - |
| $N_2H_2$ | 31.818 | 17.405 | 3.434 | -7.049 | -17.503 | 12.478 | - | 8.654 | 10.451 | 12.277 |
| $VH_2$ | 32.111 | 18.766 | 7.255 | -2.902 | -12.286 | 5.229 | 4.164 | 4.933 | 7.056 | 9.952 |
| $NVH_2$ | 33.629 | 19.169 | 7.162 | -2.905 | -12.121 | - | 3.792 | 4.065 | 6.278 | - |
| $V_2H_2$ | 42.099 | 29.718 | 18.091 | 8.320 | -1.118 | 6.125 | 6.024 | 6.677 | 9.186 | 12.028 |
| $VH_3$ | 29.702 | 15.195 | 3.601 | -6.366 | -15.478 | - | 3.978 | 4.664 | 6.977 | - |
| $N_4V$ | 44.695 | 30.313 | 16.253 | 6.443 | -2.798 | - | 5.841 | 4.061 | 6.531 | - |
| $N_2H$ | 33.204 | 18.751 | 5.282 | -7.946 | -16.825 | 10.479 | 8.306 | 7.117 | 6.169 | - |
| $N_3H$ | 34.812 | 20.408 | 6.469 | -4.317 | -14.069 | 10.252 | 8.128 | 7.550 | 7.963 | 10.491 |
| $N_3VH$ | 42.037 | 28.171 | 15.781 | 5.302 | -4.391 | 9.445 | 7.859 | 7.749 | 9.550 | 12.137 |
| $NVH_3$ | 31.965 | 17.527 | 3.512 | -6.143 | -15.393 | - | 5.535 | 3.800 | 6.425 | - |
| $N_2VH_2$ | 35.550 | 21.120 | 7.110 | -2.586 | -11.819 | - | 4.968 | 3.238 | 5.822 | - |
| $N_4VH$ | 44.656 | 30.113 | 17.504 | 1.988 | -7.175 | - | 9.026 | - | 5.461 | - |
| $NV_2H_2$ | 42.881 | 29.641 | 18.297 | 8.296 | -1.205 | - | 5.172 | 6.108 | 8.387 | - |
| Si | 36.592 | 22.126 | 8.072 | -1.582 | -10.773 | 8.364 | 6.178 | 4.404 | 7.030 | 10.119 |
| SiV | 40.995 | 29.122 | 18.113 | 8.006 | -1.576 | 3.675 | 4.082 | 5.353 | 7.526 | 10.224 |
| SiVH | 40.957 | 26.434 | 15.538 | 5.487 | -3.790 | - | 4.779 | 6.163 | 8.392 | - |
| $SiVH_2$ | 41.073 | 26.415 | 12.977 | 2.546 | -6.075 | - | 8.145 | 6.987 | 8.392 | - |
| $SiVH_3$ | 39.204 | 24.690 | 11.923 | 0.890 | -8.263 | - | 9.805 | 9.318 | 10.565 | - |
| $Si_2$ | 48.606 | 33.008 | 18.947 | 8.944 | -0.471 | - | 13.392 | 11.611 | 13.888 | 16.753 |
| $Si_2V$ | 50.188 | 36.783 | 24.858 | 14.806 | 5.322 | 9.200 | 8.075 | 8.430 | 10.658 | 13.454 |
| $Si_2VH$ | 49.220 | 34.682 | 22.952 | 12.454 | 3.180 | - | 9.359 | 9.909 | 11.691 | - |
| $Si_2VH_2$ | 48.965 | 34.425 | 20.594 | 10.593 | 1.184 | 14.747 | 12.487 | 10.936 | 13.215 | 16.086 |
| $Si_2H$ | 47.597 | 32.212 | 20.058 | 8.009 | -1.218 | - | 15.981 | 16.107 | 16.338 | - |
| SiN | 40.154 | 25.689 | 12.040 | -1.007 | -10.141 | 11.151 | 8.966 | 7.597 | 6.830 | - |
| SiNH | 37.981 | 23.506 | 9.471 | -0.544 | -9.665 | - | 10.168 | 8.413 | 10.678 | - |
| SiNV | 43.983 | 29.555 | 18.524 | 8.434 | -0.854 | - | 3.740 | 4.989 | 7.179 | - |
| SiNVH | 42.848 | 29.079 | 15.896 | 6.102 | -3.192 | - | 6.649 | 5.746 | 8.232 | - |
| $SiNVH_2$ | 43.013 | 28.504 | 15.191 | 3.613 | -5.480 | - | 9.459 | 8.426 | 9.128 | - |
| $SiNVH_3$ | 42.060 | 27.552 | 13.517 | 3.035 | -6.867 | 14.120 | 11.892 | 10.137 | 11.935 | 14.313 |
| $SiN_2V$ | 46.865 | 32.824 | 18.961 | 9.115 | -0.178 | 7.995 | 6.234 | 4.651 | 7.085 | 10.072 |
| $SiN2VH$ | 47.231 | 32.710 | 18.847 | 6.514 | -2.677 | - | 9.505 | 7.922 | 7.869 | - |
| $SiN_2VH_2$ | 45.467 | 31.012 | 16.984 | 6.204 | -4.030 | - | 11.192 | 9.444 | 10.944 | - |
| $SiN_3V$ | 49.343 | 35.362 | 22.646 | 9.454 | 0.298 | - | 7.997 | - | 6.649 | 9.773 |
| $SiN_3VH$ | 49.462 | 34.993 | 20.986 | 9.757 | 1.351 | - | 11.013 | 9.286 | 10.337 | - |
| $SiV_2$ | 54.986 | 43.131 | 31.513 | 21.248 | 10.274 | 8.574 | 8.999 | 9.661 | 11.676 | 12.982 |
| $SiV_2H$ | 49.032 | 36.992 | 25.988 | 15.930 | 6.720 | 6.005 | 6.245 | 7.521 | 9.743 | - |
| $SiV_2H_2$ | 42.918 | 33.126 | 22.523 | 12.383 | 2.770 | 3.276 | 5.764 | - | - | 12.248 |

### 3.2.1. Nitrogen doped diamond

The calculated defect equilibria for diamond crystals doped only with nitrogen are shown in Fig. 4. The equilibria are qualitatively the same for crystals containing 1 ppb to 1000 ppm nitrogen concentrations. In all cases, the most thermodynamically stable defect at a lower temperature range is $N_4V$, followed by $N_3V$ and $N_2V$, which agrees with the experimentally observed nitrogen aggregation path during heat treatments [37]. It is known from heat treatment experiments that there is a thermodynamic drive towards the formation of $N_4V$ [56]. We can also see that $N_\alpha V$ defects anneal out at higher temperatures. This observation is also in agreement with the experimental data as it is well established that NV, $N_2V$, $N_3V$, and $N_4V$ defects are annealed out at ultra-high temperatures [37, 55, 87].

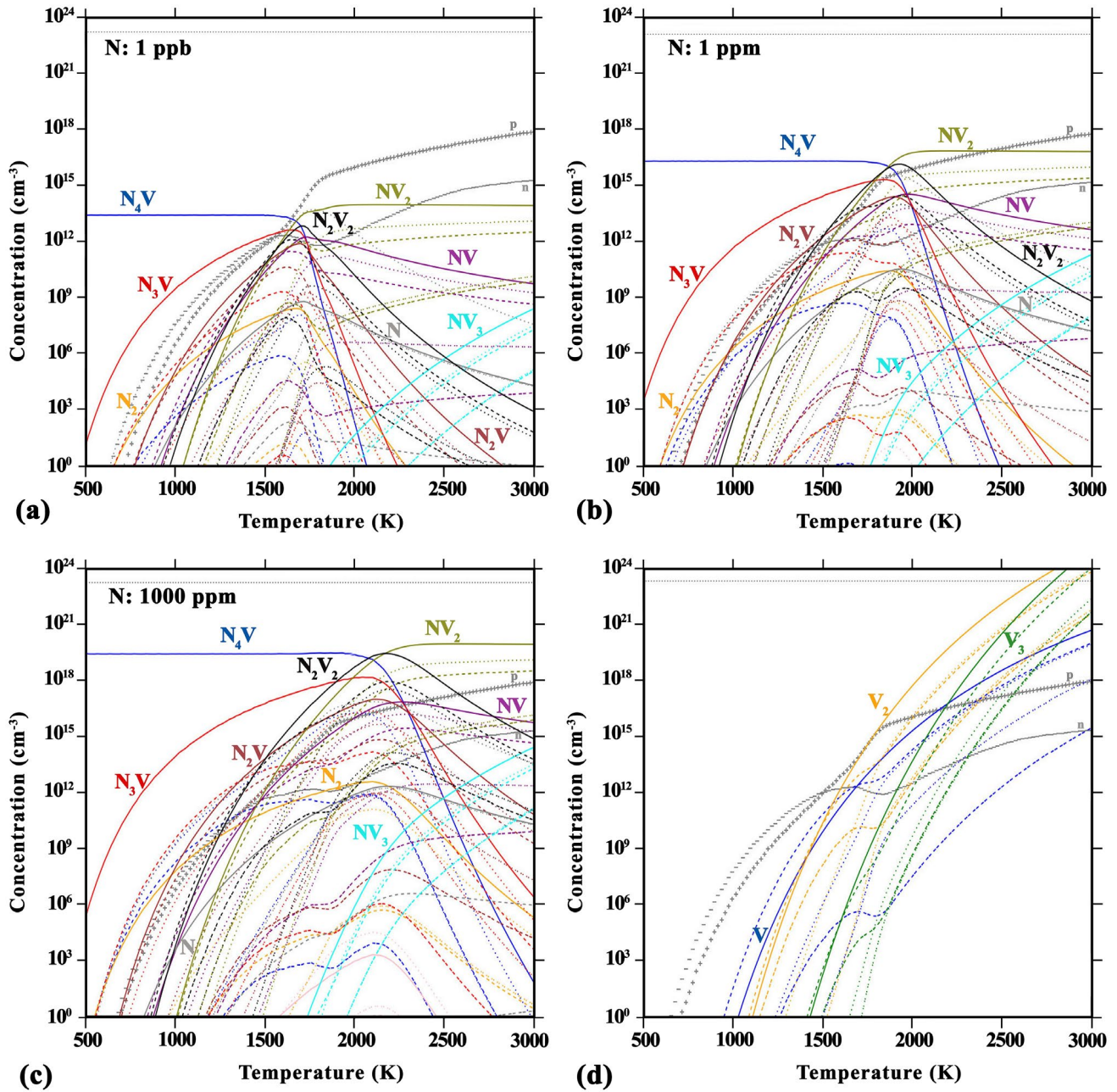

Fig. 4. Defect equilibria of $N_\alpha V_\beta$ defects in diamonds with (a) 1ppb, (b) 1ppm, (c) 1000 ppm nitrogen content demonstrates the thermodynamic drive towards formation of $N_4V$ at temperatures below 2000 K, and temperature dependence of $N_\alpha V$ defect concentrations, followed by their annealing at higher temperatures. The intrinsic defects (d) show the dominance of extended vacancy defects. Solid lines represent neutral defects, while - - (· ·) show -1 (+1) charged and -- -- (···) -2 (+2) charged defects. Free holes (p) and electrons (n) are shown by plus and minus signs respectively. $N_\alpha V_\beta$ defects present in table 2, which are not shown here are thermodynamically unstable.

A comprehensive experimental study of defect transformations during heat treatments of nitrogen doped diamond, carried out by Zaitsev et al. [88], shows that the concentration of the NV center is highly temperature dependent, with the peak intensity occurring at approximately 1000 to 1500 °C range (Fig. 5a). Our ab-initio calculations have reproduced this behavior of the NV center (Fig. 4). Zaitsev et al. [88] have reported a similar behavior for multiple nitrogen related centers, including $N_2V$, while this phenomenon is barely seen for the $N_3V$ defect in their experiments. Our calculations show that the behavior of all $N_\alpha V$ defects should be similar in this regard, mainly due to the changes in the thermodynamic activity of N with respect to temperature, as shown in Fig. 5b. We hypothesize that this phenomenon was not witnessed experimentally for the $N_3V$ defect as vividly due to the relatively short treatment times employed and the greater complexity of the $N_3V$ defect, which would require longer durations to reach equilibrium, in comparison to simpler defect complexes such as NV, as the nitrogen aggregation process is inherently diffusion-controlled. This hypothesis is based on the prior work by Koga et al. [89], where they have shown that the diffusion coefficients of more complex aggregated nitrogen related defects are orders of magnitude lower. However, Mora et al. [55] and Collins et al. [90] have reported the same temperature dependent behavior for the $N_3V$ and $N_2V$ defects, respectively. They witnessed a rise in the concentration of these defects with increasing temperature, followed by their annealing at 2500 °C and beyond. Furthermore, Zaitsev et al. [88] have observed the difference between the maximum photoluminescence intensity of various defects versus their initial condition before treatment, changes based on total nitrogen content. For diamond with lower nitrogen concentration, the change in PL intensity is more drastic by several orders of magnitude, as seen in Fig. 5a. Our computational results point in this direction as well. Considering that the as-received samples of Zaitsev et al. were HPHT treated, the as-received equilibria would be based on a temperature of about 2500 K. Considering the thermodynamic activities shown in Fig. 5b, the peak concentration must depend on total nitrogen concentration, and increasing the nitrogen content would inevitably reduce the change in peak concentrations. This phenomenon is also manifested in the equilibria shown in Fig. 4. An additional output of our theoretical approach is the nitrogen concentration dependence of the annealing temperature of the nitrogen related defect complexes. Higher nitrogen concentration in the diamond lattice increases the annealing temperature of $N_\alpha V$ defects by several hundred Kelvins (Fig. 4) because of the variation of $a_N$ with temperature and total nitrogen content (Fig. 5b). The scatter [37] in the experimental data regarding the annealing temperatures of different nitrogen related defects may be attributed to variations in the total nitrogen content in different diamond crystals.

At temperatures beyond 2000 K, the combined effect of $a_v$ and $a_N$, promotes the annealing of the $N_aV$ defects and shifts stability towards $NV_2$. Heat treatment experiments by Jackson et al. [91] have pointed towards $NV_2$ formation at temperatures above 2000 K as well. However, at extreme temperatures, over 2500 K, the emergence of extended point defects or line defects such as nitrogen platelets and voidites are widely reported [37, 51]. We have not included such extended defects in our study because their structures are unclear and subject to intense debate [92-95]. Nevertheless, the exponential increase in the concentration of multi-vacancy defects at such temperatures, as seen in Fig. 4c, can indicate the domination of extended vacancy related defects as observed in the experimental studies [51, 92, 94]. By extending the dataset in Table 2 and including the DFT calculated energies of more extended defects, one can incorporate them into the monolithic Kröger-Vink diagrams by using the approach presented in Section 2 in order to create a more comprehensive overview of defect equilibria as a function of a crystal's trace chemistry and heat treatment parameters.

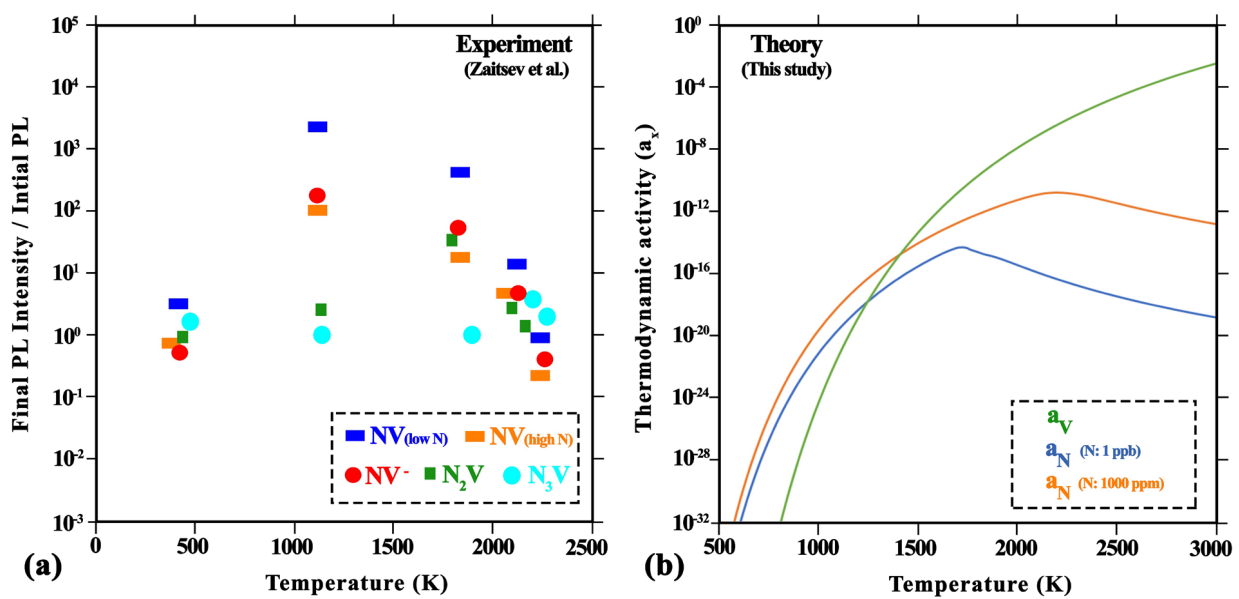

Fig. 5. The observed photoluminescence peaks before and after heat treatments (a) adapted from the prior experimental work of Zaitsev et al. [88], shows a peak concentration of $N_\alpha V$ defects at a particular temperature. The calculated thermodynamic activity of nitrogen (b) demonstrates a similar behavior and thus the main cause for the experimentally seen rise and fall of the concentrations of nitrogen related defects.

Our calculations also elucidated the partial dissociation of aggregated nitrogen related defects at a relatively limited temperature range in the vicinity of 2000 K, which gives rise to a significant concentration of positively charged, and neutral single substitutional nitrogen (X and C centers) that are annealed at higher temperatures (Fig 4). Various authors [38, 87, 96, 97] have demonstrated this phenomenon through heat treatment experiments as well. Our calculations also point towards the increase of the $N_2V_2$ defect in diamond crystals with nitrogen content over 1000 ppm, which are heat treated in a temperature range of 2000 to 2500 K (Fig. 4c). Spectroscopic data for these defects are lacking. Some of the unidentified peaks [15, 37] present in the spectroscopic measurements of a variety of heat treated diamonds may be related to these defects. Therefore, further calculations on the optical and spectroscopic properties of $N_2V_2$ and $NV_2$ defects are essential, as it is possible to make them the dominant nitrogen related defects through prolonged treatments.

Amongst the intrinsic defects, extended vacancies such as $V_2$ and $V_3$ are more dominant in concentration rather than single carbon-vacancy (Fig. 4d). A contribution of the extended vacancy defects can be their positive impact on nitrogen diffusivity. Shiryaev et al. [98] have shown that the diffusion kinetics of nitrogen increases after multiple long HPHT treatments, which supports this observation.

### 3.2.2. Nitrogen and hydrogen co-doped diamond

According to our calculations, co-doping of nitrogen and hydrogen does not affect the qualitative trends in concentrations of $N_aV_b$ defects. However, hydrogen containing defect complexes are significantly affected by co-doping and partial pressure of hydrogen.

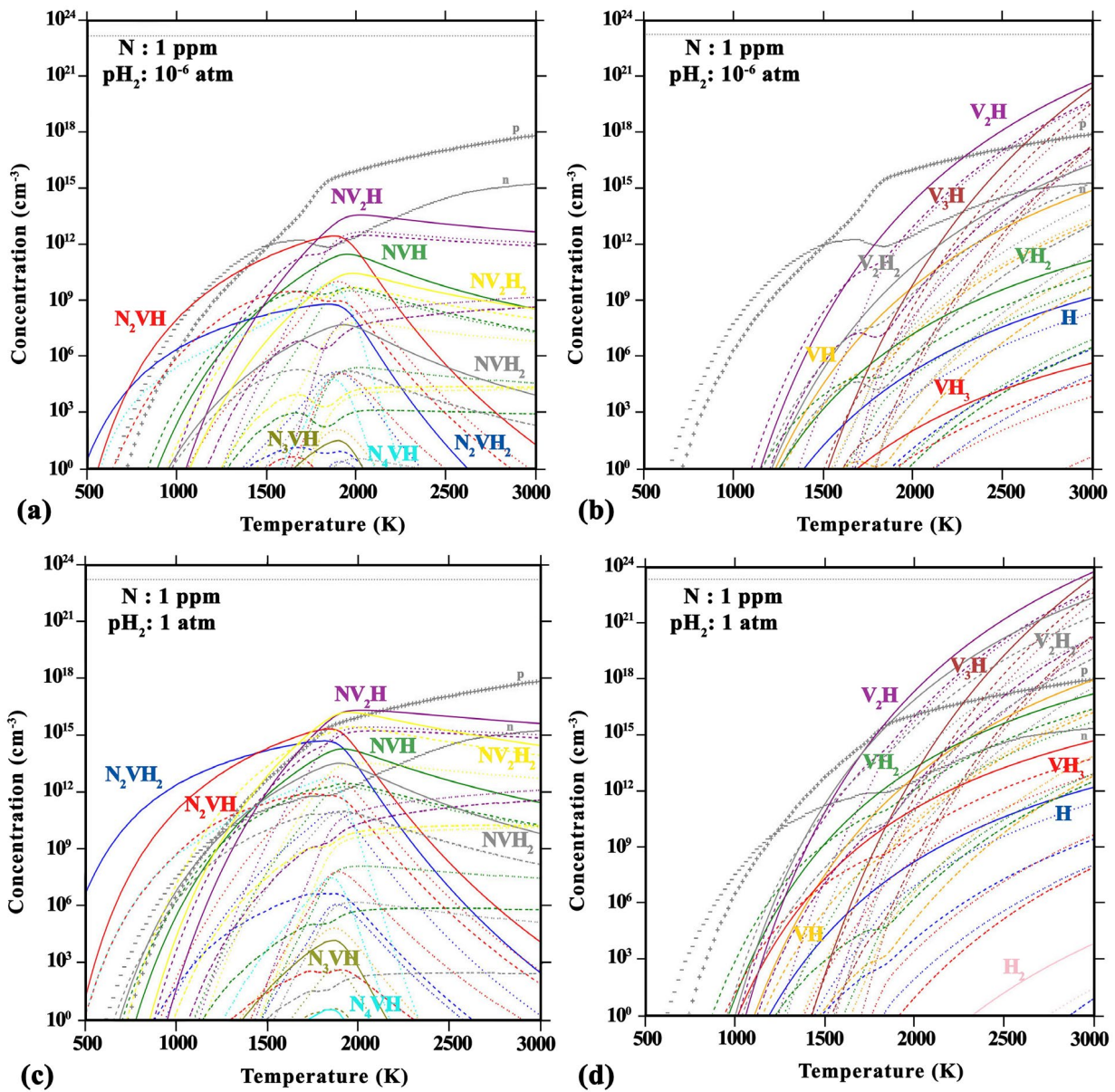

Fig. 6. Defect equilibria of $N_\alpha V_\beta H_\delta$ defects ($\alpha, \beta, \delta \geq 1$) for diamond crystals containing 1ppm nitrogen treated under $10^{-6}$ atm $pH_2$ (a) and 1 atm $pH_2$ (c), show the dominance of $N_2VH$, $N_2VH_2$ and $NV_2H$ defects. $V_\beta H_\delta$ equilibria for hydrogen poor (b) and rich conditions (d) demonstrates the dominance of $V_2H$. Solid lines represent neutral defects, while - - (· ·) show -1 (+1) charged and -- -- (·· ··) -2 (+2) charged defects. Free holes (p) and electrons (n) are shown by plus and minus signs respectively. Hydrogen containing defects in table 2, which are not shown here are thermodynamically unstable.

For diamond crystals containing 1 ppm nitrogen, treated under two different hydrogen partial pressures, $10^{-6}$ and 1 atm. the dominant hydrogen related complexes are $N_2VH$, $N_2VH_2$, $NV_2H$, $NV_2H_2$, and $V_2H$. The $N_aVH_b$ complexes anneal out at temperatures above 2000 K, while $N_aV_2H_b$ complexes are the dominant defects at the ultra high temperature range (Fig. 6). By increasing the $pH_2$ during treatment, di-hydrogen complexes exponentially rise in concentration, passivating more of the dangling bonds.

The experimental study of Hartland [99] shows that the $N_2VH$ defect is readily formed in CVD grown diamond crystals which have been heat treated at 1800 °C, while this defect's formation does not cause a reduction in the NVH concentration. Both of these observations are consistent with the calculated trends shown in Fig. 6. Hartland has also observed minute formations of $N_3VH$ defects during such treatments and hypothesized that $N_2VH$ is an intermediate step towards the formation of $N_3VH$. Therefore, they have recommended further experimental work at higher temperatures to observe the trend in the concentration of $N_3VH$ versus $N_2VH$. Our calculations show that $N_3VH$ should occur at minute concentrations at 1800 °C; however, this defect cannot dominate at the expense of $N_2VH$ as hypothesized. Based on our calculations, we expect NVH, $N_2VH$, and $N_3VH$ to anneal out at higher temperatures and pave the way for the appearance of their di-vacancy counterparts. The experimental work by Zaitsev et al. [88] also show a peak concentration for the NVH defect when treated at approximately 1800 °C, and Cruddace [100] has also experimentally shown the subsequent annealing of the NVH center (neutral and negative) at higher temperatures, both of which are consistent with the equilibria shown in Fig. 6. However, it should be noted that our calculations indicate that the temperature at which peak concentrations occur depends on the total nitrogen content, as elaborated in section 3.2.1.

There is an ongoing debate regarding the structure of an experimentally observed $V_aH_b$ type defect [101, 102]. Our results demonstrate the dominance of $V_2H$ in neutral and negative charge states (Fig. 6b), which supports the analysis of Shaw et al. [102]. By increasing the partial pressure of hydrogen to 1 atm during heat treatment (Fig. 6d), the most dominant $V_aH_b$ type defect remains to be $V_2H$; however, there is a significant rise in the concentration of $V_2H_2$ in neutral and negative charge states as well.

### 3.2.3. Nitrogen, hydrogen, and silicon co-doped diamond

In N, H, and Si co-doped diamond crystals, one should expect the formation of defect complexes comprised of all three elements. We have plotted monolithic Kröger-Vink diagrams for the N-H-V-Si complexes listed in table 2 by assuming a crystal of 1 ppm nitrogen and 1 ppm silicon content, treated under $10^{-6}$ atm hydrogen pressure (Fig. 7). Amongst the silicon containing defect complexes, the most dominant ones at temperatures below 2000 K are $SiN_2V$ in neutral state and SiNV in both neutral and -1 charge states. At higher temperatures, the equilibria shift in favor of SiV (neutral and -1 charge state), a defect readily reported in CVD diamond [103]. Furthermore, our calculations indicate that single substitutional Si and aggregated silicon complexes such as $Si_2$ should not be expected in such crystals. Therefore, our conclusions are in agreement with the prior investigations of Goss et al. [104]. The presence of the $SiN_3V$ defect (+1 charge state) is also verified in our calculations; however, the experimental validation of these defects can be rather difficult because of its extremely low-temperature stability. Among the hydrogen related complexes, the concentrations of $SiV_2H$ (in neutral and -1 charge state) and $SiV_2H_2$ (in -2 charge state) are most dominant.

When the behavior of $N_aV_bH_c$ type of defects are considered in the presence of Si, our calculations indicate that the highest concentration defects among all (including silicon containing defects) are still $N_4V$ and $NV_2$ at the lower and higher temperature range, respectively. We have not shown the $N_aV_bH_c$ defects in Fig. 7 to avoid redundancy and make the plots easier to read, as the qualitative equilibria for $N_aV_bH_c$ defects given in Fig. 4 and 6 are not considerably affected by 1 ppm of Si.

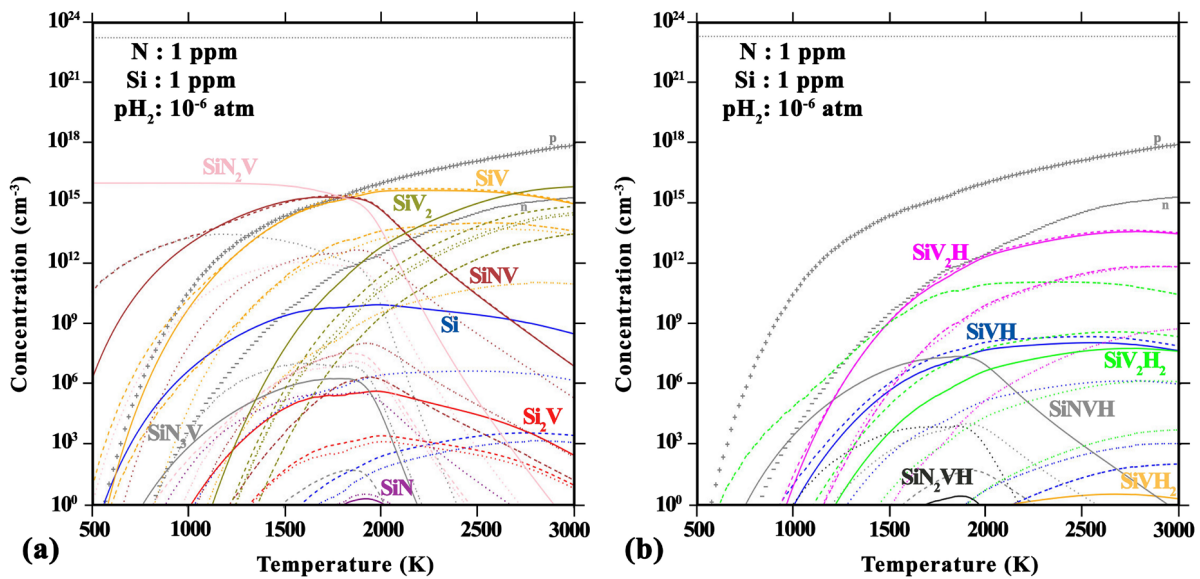

Fig. 7. Defect equilibria of $Si_\alpha N_\beta V_\delta$ defects ($\alpha, \beta, \delta \geq 1$) for diamond crystals containing 1ppm nitrogen treated under $10^{-6}$ atm $pH_2$ (a) demonstrates the dominance of $SiN_2V$, SiNV, SiV and $SiV_2$ as temperature rises. The equilibria for $Si_\alpha N_\beta V_\delta H_\lambda$ show the dominance of $SiV_2H$, and $SiV_2H_2$ defects. Solid lines represent neutral defects, while - - (· ·) show -1 (+1) charged and -- -- (···) -2 (+2) charged defects. Free holes (p) and electrons (n) are shown by plus and minus signs respectively. Silicon containing defects available in table 2, which are not shown here are unstable from a thermodynamic standpoint.

## 4. Conclusions

We have demonstrated an ab-initio pathway for modeling point defect equilibria during heat treatments of semiconducting and dielectric materials through plotting monolithic Kröger-Vink diagrams. By applying the method demonstrated in this study to nitrogen, hydrogen, and silicon doped diamond crystals, we have reproduced major experimental observations regarding nitrogen aggregation and defect transformations during heat treatments of diamond. Our key novel findings on this system are as follows.

- If nitrogen atoms can equilibrate with the surrounding nitrogen reservoir outside of the crystal, the impact of pressure on nitrogen related defects becomes significant, with the equilibria shifting in favor of the $N_4V$ defect as the pressure increases. However, this impact of pressure on defect equilibria cannot be witnessed during heat treatments of bulk diamond crystals.
- $N_4V$ is the most dominant of the $N_aV$ group of defects. However, all such defects are annealed out over approximately 2000 °C, after which $NV_2$, $VH_2$, and $V_a$ defects are thermodynamically favored.
- The exact temperature at which a nitrogen related defect anneals out depends strongly on the trace chemistry of diamond. Our calculations show that the temperature at which $N_aV$ defects anneal out can increase by 500 °C when the total nitrogen content increases from 1 ppb to 1000 ppm.
- Amongst nitrogen-hydrogen complexes, $N_2VH$ and $NV_2H$ are highly dominant at the lower and higher temperature limits, respectively, while increasing $pH_2$ during heat treatments will promote the dominance of $N_2VH_2$ and $NV_2H_2$ defects.
- In the case of co-doping with silicon, $SiN_2V$ and $SiNV$ defects are the dominant Si-N-V complexes at temperatures below 2000 Kelvin, while higher temperatures promote SiV defect's dominance in neutral and negative charge states, followed by $SiV_2$ at temperatures above 2500 Kelvin. In the presence of hydrogen, $SiV_2H$ is especially dominant.

**Future perspective**

Although the neutral defect concentrations are not influenced by the equilibrium Fermi energy for a given trace chemistry, the relative concentration of charged defects does. In our calculations, the equilibrium $E_f$ that can achieve charge neutrality at room temperature is approximately 3 eV, for 1 ppm nitrogen, which explains the significantly higher concentration of $N^+$ to $N^0$ in Fig. 4. Given that EPR or ESR cannot detect N+, its detection poses serious challenges [105]. Despite this difficulty, it is not rare to see reports on a higher concentration of positive substitutional nitrogen than its neutral counterpart in CVD grown diamond [106]. However, most experimental studies [106, 107] report their concentrations within one order of magnitude, with the dominance of the neutral charge state. Deàk et al., [49] had rightfully evaluated an equilibrium $E_f$ of 4 eV, explaining the experimentally reported relative concentrations of charged and neutral $N_s$. The defects listed in Table 2 cannot justify an $E_f$ value of 4 eV; therefore, we anticipate additional defects with shallow donor levels, which we have not considered. In other words, the defects considered in this study are by no means a complete representation of the plethora of defects, which can arise due to nitrogen, hydrogen, vacancy, and silicon in diamond. Nevertheless, despite the limited number of calculated defects, the success of this approach in capturing the major defect transformation trends in heat treatments of N, H, Si doped diamond, and the prior successful demonstration of this approach in alumina and zirconia by the Yildiz research group [60, 61], encourages us to embark on developing point defect databases. This proof of concept study shows that by creating a large enough database, which includes a wide range of dopants, it is possible to calculate the experimentally observed defect transformations during heat treatments. By acquiring a theoretical knowledge of defect equilibria as a function of trace chemistry and heat treatment process parameters, it is possible to apply state-of-the-art ab-initio methods to calculate the dominant defects' optoelectronic behavior in novel co-doped systems, minimizing the trial and error based experimental work for novel materials design.

This approach can be applied to most semiconducting and dielectric materials, where configurational entropy is more dominant than vibrational and electronic entropies. Therefore, we are in the process of developing a user-friendly open-source code for post processing of DFT results in order to automate the calculation of defect equilibria through defect databases in a wide range of materials, including nitrides and oxides. In other words, it is crucial to develop large enough databases on a considerable variety of defects through cost-effective ab-initio calculations. Although the DFT methodology used in this study can form large databases with affordable computational resources, the recent developments [108, 109] regarding on-the-fly machine learning force-fields (MLFF) can accelerate this process even further. Therefore, such an ab-initio approach can significantly accelerate defect engineering of co-doped systems and high-throughput atomic scale materials design.

**Code and data availability**

We will be happy to share the calculated data and the python code developed by our group for post processing of the DFT results, upon a reasonable request. The data and the code can be used for expanding the database, and plotting additional monolithic Kröger-Vink diagrams for miscellaneous treatment conditions or trace chemistries not considered here.

**Acknowledgments**

We are very grateful for the fruitful communications with Dr. Cristoph Freysoldt, Dr. René Windiks, Prof. Dr. Tolga Birkandan, Dr. Garip Erdoğan, Prof. Dr. Hüseyin Çimenoğlu, Prof. Dr. Burak Özkal and Prof. Dr. Servet Timur. We also appreciate the artwork provided by Wilma Van Der Giessen for our graphical abstract. The computational resources for this study have been provided by the National Center for High-Performance Computing of Turkey (UHeM), under grant number 1008852020, for which we are very thankful.

## Appendix A – Equilibrium Fermi energy

As discussed in section 2.2, the equilibrium Fermi energy of a crystal depends on the heat treatment process parameters (temperature and partial pressure of gases) and trace chemistry. By considering the defects listed in table 2, the equilibrium Fermi energy for various situations are shown in Fig. S1. The variations in equilibrium $E_f$ can be as high as 1.5 eV for a given crystal and heat treatment condition. The convergence of the $E_f$ at temperatures above 1800 K, to the value of 1.8 eV is primarily due to intrinsic charge carrier concentrations, which dominate Eq. 3.

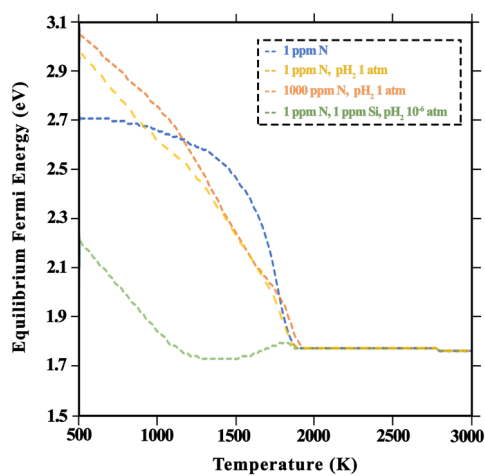

Fig. S1. Variations in equilibrium Fermi energy with respect to temperature, trace chemistry and process parameters.

## Appendix B – Benchmarking of formation energies.

The GGA calculated energies in this study are well capable of preserving the trends presented by prior HSE06 calculations of Deak et al. [49], Zemla et al. [80], and Czelej et al. [78, 79], as shown in Fig. S2. The concentrations of the neutral defects are not affected by the equilibrium Fermi energy. Therefore, considering the remarkable resemblance of the neutral defect energies between this study and prior HSE studies, as well as the capacity of the corrected GGA-PBE calculations in capturing the general trends of the formation energies for various defects, makes it possible to deduce defect equilibria with a fraction of the computational cost. Moreover, the variations in the Equilibrium Fermi energy with respect to temperature (Fig. S1) makes it less important to know the exact value of defect transition energies in the band gap. Therefore, the error of the corrected GGA-PBE results are well capable of reproducing the defect equilibria in a qualitative manner. Once the dominant defects in a system are identified, further refinements with HSE level of theory can be done. This initial screening by GGA-PBE level of theory can accelerate the process significantly.

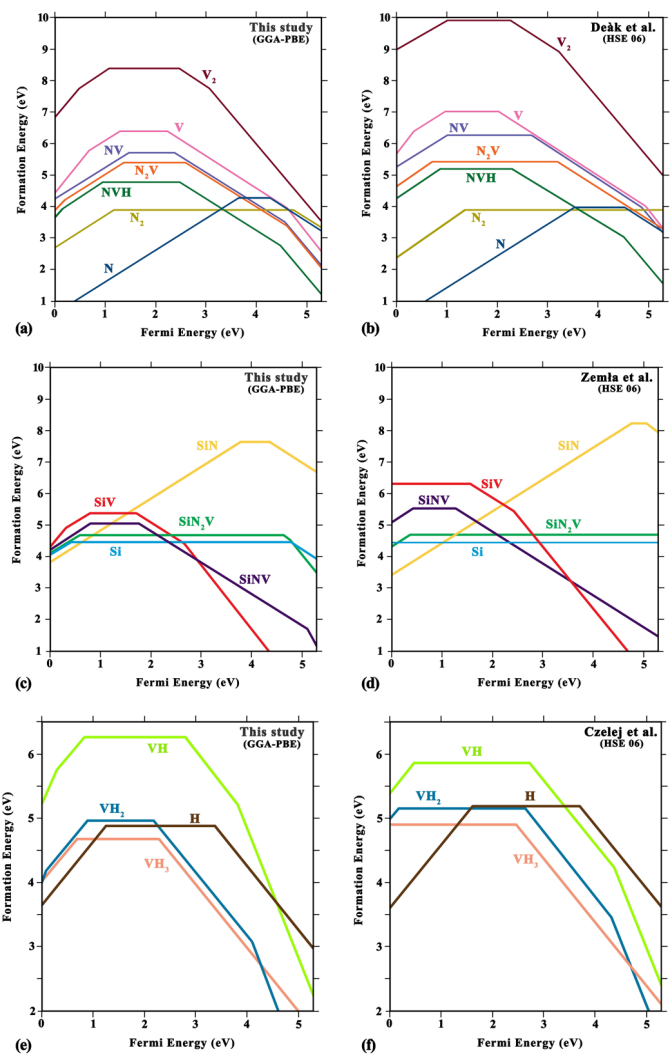

Fig. S2. Formation energy versus fermi energy for various nitrogen related and vacancy defects (a) are compared with the prior work of Deak et al. [49] (b). Similarly, the energies for silicon related defects (c) are compared with prior work by Zemla et al. [80]. Hydrogen related defects (e) are compared with the prior calculations (f) of Czelej et al. [78].